\documentclass[twocolumn,prb,showpacs]{revtex4}
\usepackage{graphicx}
\usepackage{dcolumn}
\usepackage{bm}
\usepackage{amsmath}

\begin{document}

\preprint{}
\title{Josephson and proximity effects on the surface of a topological insulator}
\author{Takehito Yokoyama}
\affiliation{Department of Physics, Tokyo Institute of Technology, Tokyo 152-8551,
Japan 
}
\date{\today}

\begin{abstract}
We investigate Josephson and proximity effects on the surface of a topological insulator on which superconductors and  a ferromagnet are deposited. The superconducting regions are described by the conventional BCS Hamiltonian, rather than the superconducting Dirac Hamiltonian. Junction interfaces are assumed to be dirty. We obtain analytical expressions of the Josephson current and the proximity-induced anomalous Green's function on the topological insulator. The dependence of the Josephson effect on the junction length, the temperature, the chemical potential and the magnetization is discussed. It is also shown that the proximity-induced pairing on the surface of a topological insulator includes even and odd frequency triplet pairings as well as a conventional $s$-wave one. 

\end{abstract}

\maketitle

\section{Introduction}

Topological insulator offers a new state of matter topologically
different from the conventional band insulator.~\cite{Moore,Fu0,Schnyder,Qi0,Hasan,Qi1} Edge channels or surface states of the topological insulator are topologically protected  and described by Dirac fermions at low energies. 
The nature of the surface Dirac fermion of the topological insulator manifests itself in interesting phenomena such as the quantized
magneto-electric effect~\cite{Qi,Qi2}, giant spin rotation~\cite{Yokoyama1},
magnetic properties of the surface state~\cite{Liu}, magnetization dynamics~\cite{Garate,Yokoyama3,Yokoyama5}, magneto-transport phenomena~\cite{Yokoyama2,Garate2,Mondal,Burkov,Yokoyama4,Culcer}, and superconducting proximity effect\cite{Fu,Fu2,Akhmerov,Tanaka,Linder}.
 
There have been a great and increasing interest on topological insulators attached to superconductors. In particular, Majorana fermions emerging in these systems have been intensively investigated.\cite{Beenakker,Alicea}  When superconductor/ferromagnet junctions are deposited on the topological insulators, surface Dirac fermions acquire a domain wall structure of the mass. At the domain wall,  Majorana fermions emerge as a zero energy bound state\cite{Fu3,Su}. Majorana fermions  have received much interest from the viewpoint of fundamental physics and also fault-tolerant  quantum computing due to their exotic properties\cite{Beenakker,Alicea}.
It has been also shown that Majorana bound states crucially influence the Josephson effect. The current-phase relation shows 4$\pi$ periodicity, i.e., $ \sin (\phi/2)$ with the phase difference across the junction $\phi$\cite{Kitaev,Kwon,Fu3}. In previous works,\cite{Beenakker,Alicea,Fu3,Tanaka,Linder}  it is assumed that the Dirac fermions become superconducting due to the proximity effect, and the junctions between superconducting and normal Dirac fermions are considered. 
In this paper, we take a different modeling of the same system. We consider the coupling between  conventional superconductors and a topological insulator, rather than that between superconducting and normal (or magnetic) Dirac fermions.\cite{Stanescu,Lababidi,Grein}  Namely, tunneling between the Schr\"odinger electrons and the Dirac fermions is explicitly taken into account.
Here, the superconductors are topologically trivial and  hence, in this setup, there appear no Majorana fermions.

In this paper, we study Josephson and proximity effects on the surface of a topological insulator on which superconductors and  a ferromagnet are deposited. The superconducting regions are described by the conventional BCS Hamiltonian, rather than superconducting Dirac electrons. We consider disordered junction interfaces in contrast to the previous works.\cite{Beenakker,Alicea,Fu3,Tanaka,Linder} We obtain analytical expressions of the Josephson current and the proximity-induced anomalous Green's function on the topological insulator. The dependence of the Josephson effect on the junction length, the temperature, the chemical potential and the magnetization is discussed. It is also shown that the proximity-induced pairing on the surface of a topological insulator includes even and odd frequency triplet pairings\cite{Berezinskii,Tanaka2} as well as a conventional $s$-wave one.  

Previous works on the Josephson effect on the surface of a topological insulator are mostly based on the approach which takes into account only the contribution from the Andreev bound states.\cite{Fu3,Tanaka,Linder} This holds for short junctions with $d  \ll \xi$, where $d$ and $\xi$ are the junction length and the superconducting coherence length, respectively. \cite{Beenakker2}
In this paper, we adopt the functional integral method\cite{Popov,Nagaosa} which is applicable to any length of the junction\cite{Awaka,Mori}, thus allowing to study the asymptotic behavior of the Josephson current for $d \to \infty$. 

\section{Formulation}

\begin{figure}[tbp]
\begin{center}
\scalebox{0.8}{
\includegraphics[width=10.0cm,clip]{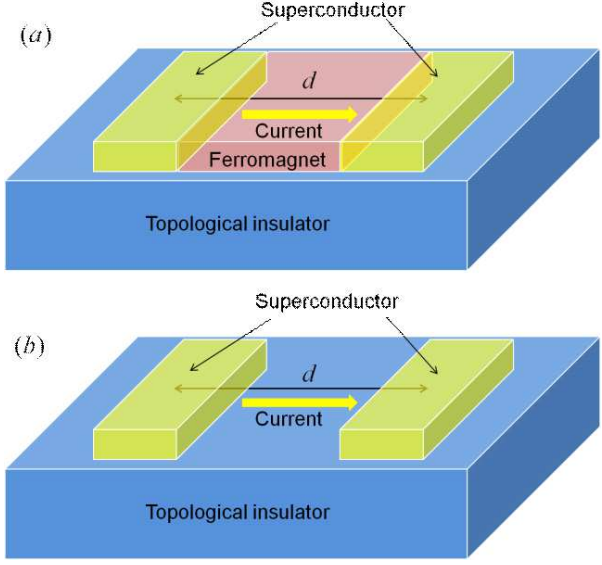}
}
\end{center}
\caption{(Color online) Schematic of the model. }
\label{fig1}
\end{figure}

We consider superconductor/topological insulator/superconductor junctions where a ferromagnet is also attached to the topological insulator, as shown in Fig. \ref{fig1} (a). Junctions without the ferromagnetic region (Fig. \ref{fig1} (b)) can be considered just by setting the exchange field to zero in the ferromagnetic region. 
The total Hamiltonian of the system reads
\begin{eqnarray}
H = H_L  + H_R  + H_M  + H_T 
\end{eqnarray}
where \cite{Fu,Mori}
\begin{widetext}
\begin{eqnarray}
 H_{L(R)}  = \sum\limits_{{\bf{k}}_{L(R)} } {\phi _{{\bf{k}}_{L(R)} }^\dag  \xi _{{\bf{k}}_{L(R)} } \sigma _0  \otimes \tau _3 \phi _{{\bf{k}}_{L(R)} }^{} } 
+ \sum\limits_{{\bf{k}}_{L(R)} } {\phi _{{\bf{k}}_{L(R)} }^\dag  \left[ {\Delta e^{- i\varphi _{L(R)} \tau _3 } \sigma _0  \otimes \tau _1 } \right]\phi _{{\bf{k}}_{L(R)} }^{} },  \\ 
 H_M  = \sum\limits_{\bf{k}} {\phi _{\bf{k}}^\dag  \left[ {\hbar v_F \left( {k_y \sigma ^x  - k_x \sigma ^y } \right) \otimes \tau _3  + {\bf{m}} \cdot {\bm{\sigma }} \otimes \tau _0  - \mu \sigma _0  \otimes \tau _3 } \right]\phi _{\bf{k}}^{} },  \\ 
 H_T  = \sum\limits_{{\bf{k}},{\bf{k}}_L} {\phi _{{\bf{k}}_L }^\dag  \left[ {te^{i({\bf{k}} - {\bf{k}}_L ) \cdot {\bf{r}}_L } \sigma _0  \otimes \tau _3 } \right]\phi _{\bf{k}}^{} }  
+ \sum\limits_{{\bf{k}},{\bf{k}}_R} {\phi _{{\bf{k}}_R }^\dag  \left[ {te^{i({\bf{k}} - {\bf{k}}_R ) \cdot {\bf{r}}_R } \sigma _0  \otimes \tau _3 } \right]\phi _{\bf{k}}^{} }  + h.c. ,
\end{eqnarray}
\end{widetext}
with $\xi _{{\bf{k}}_{L(R)} }  = \frac{{\hbar ^2 {\bf{k}}_{L(R)}^2 }}{{2m}} - \mu _{L(R)}$ and $\phi _{{\bf{k}}_{L(R)} }^\dag   = (c_{{\bf{k}}_{L(R)}  \uparrow }^\dag  , c_{{\bf{k}}_{L(R)}  \downarrow }^\dag  ,ic_{ - {\bf{k}}_{L(R)}  \downarrow }^{} , - ic_{ - {\bf{k}}_{L(R)}  \uparrow }^{} )$.
Here, $\Delta $ and $\varphi _{L(R)}$ are the magnitude of the gap function and the phase of the left (right) superconductor, respectively. Also, ${\bf{m}}$ is the exchange field, and $\sigma$ and $\tau$ are Pauli matrices in spin and Nambu spaces, respectively. 
$H_{L(R)}$ represents the Hamiltonian on the left (right) supercondutor, while $H_M $ is the Dirac Hamiltonian with the exchange field. Note that the superconductors are described by the Schr\"odinger electrons and topologically trivial. Hence, in this setup, no Majorana fermions emerge.\cite{note}
$H_T$ is the tunneling Hamiltonian between the superconductors and the surface of the topological insulator which is treated as a perturbation. 
${\bf{r}}_{L(R)}$ is the position of the interface between the left (right) superconductor and the topological insulator.
We consider the incoherent tunneling model where the spin is conserved but the momentum is not conserved upon tunneling at the interface. 
This modeling is applicable to junctions with imperfect dirty insulating barriers. In real space representation, the tunneling matrix element reads $t \delta({\bf{r}}-{\bf{r}}_{L(R)})$. 
 The average of the position vectors is assumed to give $\left\langle {\left| {{\bf{r}}_R  - {\bf{r}}_L } \right|} \right\rangle  = d$.\cite{Mori} The calculated results are averaged over the positions of ${\bf{r}}_L$ and ${\bf{r}}_R$ at the interfaces.

The partition function is then given by 
\begin{eqnarray}
Z = \int {D\bar \psi D\psi \exp \left[ - {\sum\limits_{\{ {\bf{k}}\} _{} } {\bar \psi \left( { - G_0^{ - 1}  + \hat{T}} \right)\psi } } \right]} 
\end{eqnarray}
where $\bar \psi  = \left( {\bar \phi _{{\bf{k}}_L }, \bar \phi_{{\bf{k}}}, \bar \phi _{{\bf{k}}_R }} \right)$. $G_0$ is the bulk Green's function while $\hat{T}$ is a tunneling matrix. See the Appendix for their explicit forms.
The free energy of the system can be calculated as  $F =  - T\ln Z$ where $T$ is the temperature of the system. The leading contribution to the Josephson current is given by the fourth order with respect to the tunneling Hamiltonian (see the Appendix for the details of the calculation).
The Josephson current is then calculated as 
\begin{widetext}
\begin{eqnarray}
 I =  - \frac{{2e}}{\hbar }\frac{{\partial F}}{{\partial \varphi }}
= - \frac{{4e}}{\hbar } Tt^4 \sin (\varphi  + 2m_y d/\hbar v_F  )\sum\limits_{\omega _n } {\frac{{(\nu V\Delta )^2 }}{{\omega _n^2  + \Delta ^2 }}\left[ {\left| {\hbar v_F k_F^{} } \right|^2 \left| {K_1 (k_F^{} d)} \right|^2  - (\omega _n^2  + \mu ^2  - m_z^2 )\left| {K_0 (k_F^{} d)} \right|^2 } \right]} 
\end{eqnarray}
where $\nu$, $V$, $\omega _n$, and $K_\nu(z) (\nu=0, 1)$ are, respectively, the density of states at the Fermi level, the area of the surface of the topological insulator sandwiched between the superconductors, the fermionic Matsubara frequency, and the modified Bessel function. Also, $k_F$ is defined by $\hbar v_F k_F^{}  = \sqrt {(\omega _n^{}  - i\mu )^2  + m_z^2 }$ and the branch is taken so that ${\mathop{\rm Re}\nolimits} k_F^{}  > 0$. Here, $\varphi  = \varphi _R  - \varphi _L $ is the phase difference across the junction.
It is seen that the Josephson effect is independent of $m_x$, and $m_y$ shifts the phase difference. \cite{Tanaka,Linder} 

The critical current $I_C$ can be written as
\begin{eqnarray}
 \frac{{ - eI_C R}}{{T_C }} = \frac{T}{{T_C }}\left( {\frac{d}{{\hbar v_F }}} \right)^2 \sum\limits_{\omega _n } {\frac{{\Delta ^2 }}{{\omega _n^2  + \Delta ^2 }}\left[ {\left| {\hbar v_F k_F^{} } \right|^2 \left| {K_1 (k_F^{} d)} \right|^2  - (\omega _n^2  + \mu ^2  - m_z^2 )\left| {K_0 (k_F^{} d)} \right|^2 } \right]}  \label{Ic}
\end{eqnarray}
\end{widetext}
where $T_C$ is the superconducting transition temperature and 
\begin{eqnarray}
R^{ - 1}  = \frac{{4e^2 }}{\hbar }\left( {\frac{{t^2 V\nu }}{d}} \right)^2 .
\end{eqnarray}

\section{Results}
\subsection{Josephson effect}

\begin{figure}[tbp]
\begin{center}
\scalebox{0.8}{
\includegraphics[width=10cm,clip]{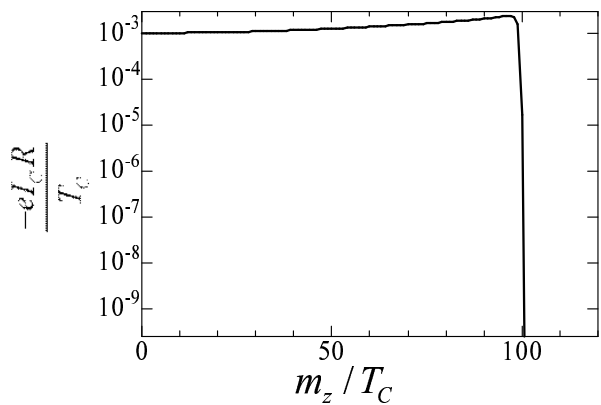}
}
\end{center}
\caption{ Critical Josephson current as a function of $m_z$ for $T/T_C=0.1$, $d/\xi=1$ and $\mu/T_C=100$. }
\label{fig2}
\end{figure}

\begin{figure}[tbp]
\begin{center}
\scalebox{0.8}{
\includegraphics[width=10cm,clip]{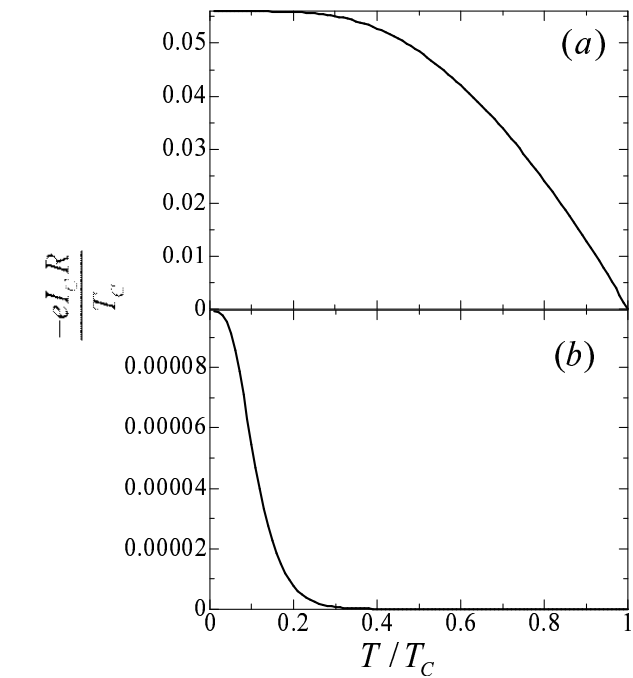}
}
\end{center}
\caption{ Critical Josephson current as a function of temperature of the system  with $m_z=0$ and $\mu/T_C=100$  for (a) $d/\xi=0.1$ and (b) $d/\xi=5$. }
\label{fig3}
\end{figure}

\begin{figure}[tbp]
\begin{center}
\scalebox{0.8}{
\includegraphics[width=10cm,clip]{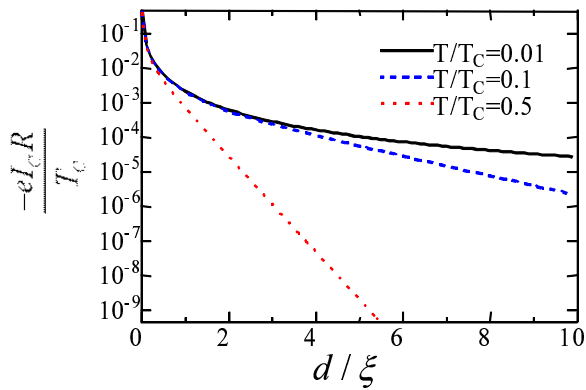}
}
\end{center}
\caption{(Color online) Critical Josephson current as a function of the distance between the superconductors $d$ for $m_z=0$,  $\mu/T_C=100$ and several temperatures. }
\label{fig4}
\end{figure}

\begin{figure}[tbp]
\begin{center}
\scalebox{0.8}{
\includegraphics[width=10cm,clip]{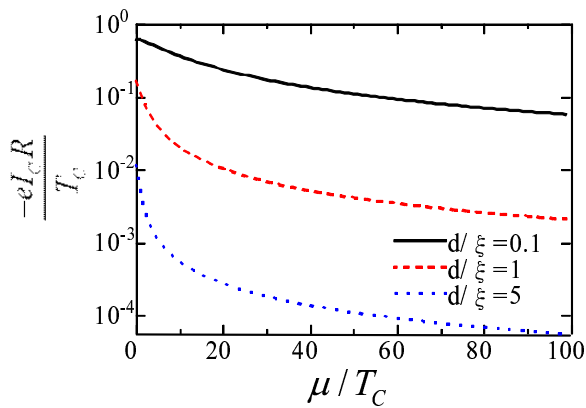}
}
\end{center}
\caption{(Color online) Critical Josephson current as a function of the chemical potential $\mu$ for $m_z=0$, $T/T_C=0.1$ and several $d$. }
\label{fig5}
\end{figure}
In what follows, we will study the critical Josephson current using Eq.(\ref{Ic}). We consider a temperature dependence of the gap of the BCS type modeled by \cite{Muhlschlegel}
\begin{eqnarray}
\Delta (T) = \Delta (0) \tanh \left( {1.74\sqrt {T_C /T - 1} } \right).
\end{eqnarray}

\subsubsection{The effect of the exchange field}
Here, let us study the effect of the exchange field.
As seen from Eq.(\ref{Ic}), the Josephson effect is independent of $m_x$, and $m_y$ shifts the phase difference.
 Since the inplane exchange field corresponds to the shift of the momentum,\cite{Yokoyama2} the effect of the inplane exchange field can be reduced to the phase factor (which can be seen by proper transformations in Eq. (\ref{f1})) and hence we find the phase shift proportional to $m_y$.

To see the effect of the $z$-component of the exchange field, $m_z$, we plot the dependence of the critical Josephson current on $m_z$  in Fig. \ref{fig2} for $T/T_C=0.1$, $d/\xi=1$ and $\mu/T_C=100$ where $\xi=\hbar v_F/T_C$ is the superconducting coherence length. 
With increasing $m_z$, $I_C$ increases and for $m_z > \mu$, the current is strongly suppressed. This is because for $m_z > \mu$, the Fermi level lies inside the mass gap and hence the surface state of the topological insulator is no more metallic. Note that the the surface spectrum is given by $E_M  =  \pm \sqrt {k^2  + m_z^2 }  \pm \mu $ (here we set $m_x=m_y=0$ for simplicity).
Also, it should be noted that since the superconductors are topologically trivial, there is no edge state, or Majorana bound state. This is in stark contrast to similar junctions where superconducting regions are described by Dirac fermions: there, the Majorana bound state can carry the Josephson current.\cite{Fu3,Tanaka,Linder}

\subsubsection{Superconductor/topological insulator/superconductor junction}
Now, we will investigate the Josephson junction characteristics. Since the effects of the exchange field have been clarified, here let us consider the junction with $\bf{m}=0$. This corresponds to the junction illustrated in Fig. \ref{fig1} (b). 

Figure \ref{fig3} depicts $T$ dependence of the critical Josephson current with $\mu/T_C=100$  for (a) $d/\xi=0.1$ and (b) $d/\xi=5$. 
For short normal segment $d/\xi=0.1$, the behavior is similar to that of the conventional Josephson junctions through an insulating barrier, i.e., $\tanh(\Delta/2T)$\cite{Ambegaokar}.
For large $d$, the critical current shows an exponential decay. 
This can be also obtained as follows. 
Using the asymptotic form of the modified Bessel function for $\left| z \right| \gg 1$:
\begin{eqnarray}
K_\nu  (z) \sim \sqrt {\frac{\pi }{{2z}}} e^{ - z} \left[ {1 + \frac{{(4\nu ^2  - 1)}}{{8z}} + ...} \right],
\end{eqnarray}
for $m_z  = 0, Td/\hbar v_F  \gg 1$ and $ \mu  \gg T,$ we have 
\begin{eqnarray}
\frac{{ - eI_C R}}{{T_C }} \sim \frac{1}{{2 \mu T_C }}\frac{{\left( {\pi T\Delta } \right)^2 }}{{\left( {\pi T} \right)^2  + \Delta ^2 }}e^{ - 2\pi Td/\hbar v_F}. \label{ica}
\end{eqnarray}
Notice that the $n=0$ component in the Matsubara frequencies has a dominant contribution to the Josephson current. 
This form shows a typical exponential decay of the critical Josephson current for $Td/\hbar v_F  \gg 1$: 
 the asymptotic behavior of the Josephson current (the exponential decay) has the same form as that governed by the Schr\"odinger electrons.\cite{Mori,Golubov}
 
In Fig. \ref{fig4}, we show $d$ dependence of the critical Josephson current for  $\mu/T_C=100$ and several temperatures. 
For large $d$ or at high temperature, we see  an exponential decay of the critical Josephson current. This is also consistent with the above analytical expression.

Figure \ref{fig5} shows $\mu$ dependence of the critical Josephson current for $T/T_C=0.1$ and several $d$. It is found that with the increase of $\mu$, the critical current monotonically decreases. This is because the proximity effect is suppressed by increasing the chemical potential $\mu$ as will be shown in  Eq. (\ref{fmh0}).  
For large $d$, the Josephson current is inversely proportional to the chemical potential as seen from Eq. (\ref{ica}).
Experimentally, the chemical potential can be tuned by chemical doping\cite{Hsieh} or gating\cite{Chen}.

Recently, Josephson supercurrent through a topological insulator surface state has been observed.\cite{Veldhorst} The dependence on $T$ and $d$ shown in Figs. \ref{fig3} and \ref{fig4} is qualitatively consistent with the experimental data. A quantitative difference would come from the fact that the bulk states of the topological insulator also contribute to the Josephson current because the chemical potential of the sample used in Ref. \onlinecite{Veldhorst} probably crosses the bulk bands.

\subsection{Proximity effect}
In this subsection, we will investigate the proximity effect in a topological insulator/$s$-wave superconductor junction. 
Proximity effect in this junction has been investigated in Refs. \onlinecite{Stanescu,Lababidi,Grein}. The tunneling between the superconductor and a bulk topological insulator has been considered, and the validity of the Fu-Kane model has been discussed, based on mostly numerical approaches.\cite{Stanescu,Lababidi,Grein} 

Here, we consider the tunneling between the superconductor and the \textit{surface} of the topological insulator.
Using the functional integral method, we derive analytical expressions of the proximity-induced anomalous Green's functions on the topological insulator.

Now, let us consider superconductor/topological insulator bilayer junctions. To do so, let us remove the degree of freedom of the right superconductor from the above formulation (see also the Appendix). Without loss of generality, we can set $\varphi_L=0$ and ${\bf{r}}_L =\bf{0}$.
The Green functions can be calculated as 
\begin{eqnarray}
G = \int {D\bar \psi D\psi \psi \bar \psi \exp \left[ -{\sum\limits_{\{ {\bf{k}}\} _{} } {\bar \psi \left( { - G_0^{ - 1}  + \hat{T}} \right)\psi } } \right]} 
\end{eqnarray}
where $\bar \psi  = \left( {\bar \phi _{{\bf{k}}_L }^{} , \bar \phi _{{\bf{k}}_{} }^{} } \right)$. Performing the functional integral, we have
\begin{eqnarray}
G = (G_0^{ - 1}  - \hat{T})^{ - 1}  = G_0^{} \sum\limits_n {(\hat{T}G_0^{} )^n } 
\end{eqnarray}
where
\begin{widetext}
\begin{eqnarray}
G_{}^{}  = \left( {\begin{array}{*{20}c}
   {G_L^{\prime} } & 0  \\
   0 & {G_M^{\prime} }  \\
\end{array}} \right),G_0^{}  = \left( {\begin{array}{*{20}c}
   {G_L^{} } & 0  \\
   0 & {G_M^{} }  \\
\end{array}} \right),\hat{T} = \left( {\begin{array}{*{20}c}
   0 & {T_{12} }  \\
   {T_{21} } & 0  \\
\end{array}} \right).
\end{eqnarray}
The leading contribution is given by the second order with respect to the tunneling matrix. The anomalous Green's function on the surface of the topological insulator $f'_M $ in the second order in $t$ can be represented as 
\begin{eqnarray}
f'_M  =  - t^2 g_M^{} ({\bf{k}},\omega _n )\bar g_M^{} ({\bf{k}},\omega _n )\sum\limits_{{\bf{k}}_L } {f_L^{} ({\bf{k}}_L ,\omega _n )}  \\ 
 = \frac{{\pi \nu \Delta t^2 }}{{\sqrt {\omega _n^2  + \Delta ^2 } }} \times \nonumber \\  \frac{{ - \left[ {\omega _n^2  + \mu ^2  + (\hbar v_F k)^2 } \right] + m^2  - 2\mu \hbar v_F {\bf{k}}_ \bot   \cdot {\bm{\sigma }} + 2i\omega _n {\bf{m}} \cdot {\bm{\sigma }} + 2i\hbar v_F ({\bf{k}}_ \bot   \times {\bf{m}}) \cdot {\bm{\sigma }}}}{{\left[ {(i\omega _n  + \mu )^2  - (\hbar v_F )^2 \left\{ {(k_y  + m_x )^2  + (k_x  - m_y )^2 } \right\}  - m_z^2  } \right]\left[ {(i\omega _n  - \mu )^2  - (\hbar v_F )^2 \left\{ {(k_y  - m_x )^2  + (k_x  + m_y )^2 } \right\}  - m_z^2 } \right]}}
\end{eqnarray}
with ${\bf{k}}_ \bot   = (k_y , - k_x ,0)$. 
Here, spin-singlet pairing is proportional to the unit matrix in spin space while spin-triplet pairing is proportional to the Pauli matrix ${\bm{\sigma }}$.
Therefore, it is seen that both singlet and triplet pairings are induced on the surface of the topological insulator. The generation of the triplet pairing reflects the symmetry breaking in spin space. \cite{Bergeret}
In particular, for ${\bf{m}}=\bf{0}$, we have 
\begin{eqnarray}
f'_M  = \frac{{\pi \nu \Delta t^2 }}{{\sqrt {\omega _n^2  + \Delta ^2 } }} \frac{{ - \left[ {\omega _n^2  + \mu ^2  + (\hbar v_F k)^2 } \right] - 2\mu \hbar v_F {\bf{k}}_ \bot   \cdot {\bm{\sigma }}}}{{(\omega _n^2  + \mu ^2 )^2  + (\hbar v_F k)^4  + 2(\hbar v_F k)^2 (\omega _n^2  - \mu ^2 )}}. \label{fmh0}
\end{eqnarray}
We see that in the limit of $\mu \to \infty$, we have $f'_M \to 0$. This explains the suppression of the Josephson curernt with $\mu$ in Fig. \ref{fig5}.
It is also found that even in the absence of the exchange field, triplet pairing is induced on the surface if $\mu \ne 0$, which is consistent with Refs. \onlinecite{Stanescu, Lababidi} (see also Ref. \onlinecite{Gorkov}). 
In previous works, it is assumed that by attaching an $s$-wave superconductor to a topological insulator, the same $s$-wave superconductivity is induced on the surface\cite{Beenakker,Alicea,Fu3,Tanaka,Linder}. Here, we find that not only $s$-wave singlet superconductivity but, in general, triplet $p$-wave superconductivity is also induced on the surface of the topological insulator. \cite{Stanescu, Lababidi}
Also, we here assume clean surface states on the topological insulator. If the surface is in the diffusive regime, it is expected that odd frequency triplet $s$-wave superconductivity is induced on the topological insulator. \cite{Asano}

Let us focus on the case with $\mu=0$ but finite exchange field. The anomalous Green's function then becomes 
\begin{eqnarray}
f'_M  = F(k_x ,k_y ,\omega _n^{} )\left[ { - \omega _n^2  - (\hbar v_F k)^2 + m^2  + 2i\omega _n {\bf{m}} \cdot {\bm{\sigma }} + 2i\hbar v_F ({\bf{k}}_ \bot   \times {\bf{m}}) \cdot {\bm{\sigma }}} \right] , \\
F(k_x ,k_y ,\omega _n^{} ) = \frac{{\pi \nu \Delta t^2 }}{{\sqrt {\omega _n^2  + \Delta ^2 } }}  \times \nonumber \\  \frac{1}{{\left[ {\omega _n^2  + (\hbar v_F )^2 \left\{ {(k_y  + m_x )^2  + (k_x  - m_y )^2 } \right\}  + m_z^2 } \right]\left[ {\omega _n^2  + (\hbar v_F )^2 \left\{ {(k_y  - m_x )^2  + (k_x  + m_y )^2 } \right\} + m_z^2 } \right]}}.
\end{eqnarray}
\end{widetext}
Note that $F(k_x ,k_y ,\omega _n^{} )$ is an even function of ${\bf{k}} (= (k_x ,k_y ,0))$ and $\omega _n^{}$. We find that the component proportional to $- \omega _n^2  - (\hbar v_F k)^2 + m^2$ represents a singlet $s$-wave superconductivity while that proportional to $2i\omega _n {\bf{m}} \cdot {\bm{\sigma }}$ is triplet and odd in $\omega _n$, namely odd frequency triplet $s$-wave pairing\cite{Berezinskii,Tanaka2}. The component proportional to $2i({\bf{k}}_ \bot  \times {\bf{m}}) \cdot {\bm{\sigma }} $ corresponds to  a triplet $p$-wave superconductivity.

\section{Summary}

In this paper,
 we have investigated Josephson and proximity effects on the surface of a topological insulator on which superconductors and  a ferromagnet are deposited. We have described the superconducting regions by the conventional BCS Hamiltonian, rather than the superconducting Dirac Hamiltonian. We have presented analytical expressions of the Josephson current and the proximity-induced anomalous Green's function on the topological insulator. The dependence of the Josephson effect on the junction length, the temperature, the chemical potential and the magnetization has been discussed. It has been also shown that the proximity-induced pairing on the surface of a topological insulator includes even and odd frequency triplet pairings as well as a conventional $s$-wave one.

This work was supported by Grant-in-Aid for Young Scientists (B) (No. 23740236) and the "Topological Quantum Phenomena" (No. 23103505) Grant-in Aid for Scientific Research on Innovative Areas from the Ministry of Education, Culture, Sports, Science and Technology (MEXT) of Japan.

\appendix

\begin{widetext}

\section{Calculation of the free energy}
Here, we present the details of the calculation of the free energy of the junctions. 
The unperturbed Green's function $G_0$ is represented by a 12$\times$12 matrix as
\begin{eqnarray}
 G_0^{ - 1}  = \left( {\begin{array}{*{20}c}
   {G_L^{ - 1} } & 0 & 0  \\
   0 & {G_M^{ - 1} } & 0  \\
   0 & 0 & {G_R^{ - 1} }  \\
\end{array}} \right), \\ 
 G_{L(R)}^{}  =  - \frac{{i\omega _n \sigma _0  \otimes \tau _0  + \xi _{{\bf{k}}_{L(R)} } \sigma _0  \otimes \tau _3  + \Delta e^{ - i\varphi _{L(R)} \tau _3 } \sigma _0  \otimes \tau _1 }}{{\omega _n^2  + \xi _{{\bf{k}}_{L(R)} }^2  + \Delta ^2 }}  \equiv \left( {\begin{array}{*{20}c}
   {g_{L(R)}^{} } & {f_{L(R)}^{} }  \\
   {\bar f_{L(R)}^{} } & {\bar g_{L(R)}^{} }  \\
\end{array}} \right)
, \\ 
G_M^{}  = \left( {\begin{array}{*{20}c}
   {\frac{{i\omega _n  + \mu  + \hbar v_F (k_y  + m_x )\sigma ^x  - \hbar v_F (k_x  - m_y )\sigma ^y  + m_z \sigma ^z }}{{(i\omega _n  + \mu )^2  - (\hbar v_F )^2 \left\{ {(k_y  + m_x )^2  + (k_x  - m_y )^2 } \right\} - m_z^2 }}} & 0  \\
   0 & {\frac{{i\omega _n  - \mu  - \hbar v_F (k_y  - m_x )\sigma ^x  + \hbar v_F (k_x  + m_y )\sigma ^y  + m_z \sigma ^z }}{{(i\omega _n  - \mu )^2  - (\hbar v_F )^2 \left\{ {(k_y  - m_x )^2  + (k_x  + m_y )^2 } \right\} - m_z^2 }}}  \\
\end{array}} \right)
 \equiv \left( {\begin{array}{*{20}c}
   {g_M^{} } & 0  \\
   0 & {\bar g_M^{} }  \\
\end{array}} \right).
\end{eqnarray}

By performing the functional integral, we have the free energy of the junctions of the form
\begin{eqnarray}
F =  - T\ln Z =  - T {\rm{Tr}}\ln \left[ { - G_0^{ - 1}  + \hat{T} } \right]
\end{eqnarray}
where $\hat{T}$ is the tunneling matrix given by
\begin{eqnarray}
\hat{T} = \left( {\begin{array}{*{20}c}
   0 & {T_{12} } & 0  \\
   {T_{21} } & 0 & {T_{23} }  \\
   0 & {T_{32} } & 0  \\
\end{array}} \right)
\end{eqnarray}
with $T_{12}  = te^{i({\bf{k}} - {\bf{k}}_L ) \cdot {\bf{r}}_L } \sigma _0  \otimes \tau _3  = T_{21}^* $ and $T_{23}  = te^{i({\bf{k}}_R  - {\bf{k}}) \cdot {\bf{r}}_R } \sigma _0  \otimes \tau _3  = T_{32}^* $.

The leading contribution is given by the fourth order with respect to the tunneling element, which is calculated as 
\begin{eqnarray}
F \approx  - \frac{T}{4}{\rm{Tr}}\left( {G_0 \hat{T}} \right)^4  \\ 
  =  - T {\rm{Tr}} \sum\limits_{{\bf{k}}_L ,{\bf{k}},{\bf{k}}_R ,{\bf{k}}',\omega _n } {G_L^{} ({\bf{k}}_L ,\omega _n )T_{12} G_M^{} ({\bf{k}},\omega _n )T_{23} G_R^{} ({\bf{k}}_R ,\omega _n )T_{32} G_M^{} ({\bf{k}}',\omega _n )T_{21} }  \\ 
  = 2Tt^4 {\rm{Tr}} {\mathop{\rm Re}\nolimits} \left[ {\sum\limits_{{\bf{k}}_L ,{\bf{k}},{\bf{k}}_R ,{\bf{k}}',\omega _n } {e^{i({\bf{k}} - {\bf{k}}') \cdot ({\bf{r}}_R  - {\bf{r}}_L )} } f_L^{} ({\bf{k}}_L ,\omega _n )\bar g_M^{} ({\bf{k}},\omega _n )\bar f_R^{} ({\bf{k}}_L ,\omega _n )g_M^{} ({\bf{k}}',\omega _n )} \right] \label{f1} \\ 
  =  - 2Tt^4 \cos (\varphi  + 2m_y d/\hbar v_F )\sum\limits_{\omega _n } {\frac{{(\nu V\Delta )^2 }}{{\omega _n^2  + \Delta ^2 }}\left[ {\left| {\hbar v_F k_F^{} } \right|^2 \left| {K_1 (k_F^{} d)} \right|^2  - (\omega _n^2  + \mu ^2  - m_z^2 )\left| {K_0 (k_F^{} d)} \right|^2 } \right]}.
\end{eqnarray}
Here, we have used the following relations
\begin{eqnarray}
J_0 (x) = \frac{1}{{2\pi }}\int_0^{2\pi } {e^{ix\cos \varphi } d\varphi } , \; K_0 (a k) = \int_0^\infty  {\frac{{xJ_0 (a x)}}{{x^2  + k^2 }}} dx , \; K_1 (x) =  - \frac{d}{{dx}}K_0 (x) 
\end{eqnarray}
for $a > 0$ and ${\mathop{\rm Re}\nolimits} k > 0$, where $J_0 (x)$ is the Bessel function.




\end{widetext}


\begin{thebibliography}{99}
\bibitem{note} In Ref. \onlinecite{Fu}, it is assumed that a finite gap function is induced on the surface of a topological insulator. On the other hand, in this paper, we assume that the suface state of a topological insulator does not acquire a finite gap function, although a superconducting correlation is induced.
\end{thebibliography}
\end{document}